# Life Beyond the Solar System: Space Weather and Its Impact on Habitable Worlds

*Airapetian, V. S.[1,2], Danchi, W. C.[1], Dong, C. F.[3], Rugheimer, S.[4], Mlynczak, M.[5], Stevenson, K. B.[6], Henning, W. G.[1,7], Grenfell, J. L.[8], Jin, M.[9], Glocer, A.[1], Gronoff, G.[5,10], Lynch, B.[11], Johnstone, C.[12], Lüftinger, T.[12], Güdel,M.[12], Kobayashi, K.[13], Fahrenbach, A.[14], Hallinan, G.[15], Stamenkovic, V.[16], Cohen, O.[17], Kuang, W.[1], van der Holst, B[18], Manchester, C.[18], Zank, G.[19], Verkhoglyadova, O.[16], Sojka, J.[19], Maehara, H.[20], Notsu, Y.[21], Yamashiki, Y.[22], France, K.[23], Lopez Puertas, M.[24], Funke, B.[24], Jackman, C.[1], Kay, C[1], Leisawitz, D.[1], Alexander, D.[25]*

[1] *NASA Goddard Space Flight Center, MD* [2] *American University, DC* [3] *Princeton University, NJ*, [4] *University of St. Andrews*, [5] *NASA Langley Research Center, VA*, [6] *Space Telescope Science Institute, MD*, [7] *University of Maryland, MD*, [8] *German Aerospace Centre, Berlin*, [9] *UCAR/LMRC*, [10] *SSAI, VA*, [11] *SSL/UC Berkley, CA*, [12] *University of Vienna, Austria*, [13] *Yokohama National University, Japan*, [14] *ELSI/Tokyo Tech, Japan*, [15] *California Institute of Technology, Pasadena, CA*, [16] *JPL/California Institute of Technology, Pasadena, CA*, [17] *University of Massachusetts, Lowell, MA* [18] *University of Michigan, Ann Harbor, MI*, [18] *University of Alabama, Huntsville, AL*, [19] *University of Utah, UT*, [20] *National Astronomical Observatory of Japan, Japan*, [20] *Kyoto University, Japan*, [23] *University of Colorado at Boulder, CO*, [24] *Institute de Astrofisica de Andalucía, Spain*, [25] *Rice University, TX.*

*Submitted to the National Academy of Sciences in support of the Astrobiology Science Strategy for the Search for Life in the Universe*

**NASA's Nexus for Exoplanet System Science (NExSS) is a research coordination network dedicated to the study of planetary habitability using a system science approach with inputs from astrophysics, Earth science, planetary science, and heliophysics. Herein, the NExSS community describes recent progress and future prospects for characterization and modeling of exoplanetary systems and technology development required to detect and identify signs of life.**

**Introduction**.

The search of life in the Universe is a fundamental problem of astrobiology and a major priority for NASA. A key area of major progress since the NASA Astrobiology Strategy 2015 (NAS15) has been a shift from the exoplanet discovery phase to a phase of characterization and modeling of the physics and chemistry of exoplanetary atmospheres, and the development of observational strategies for the search for life in the Universe by combining expertise from four NASA science disciplines (heliophysics, astrophysics, planetary science and Earth science, HAPE community). The NASA Nexus for Exoplanetary System Science (NExSS) has provided an efficient environment for such interdisciplinary studies.

Solar flares, coronal mass ejections (CMEs) and solar energetic particles (SEPs) produce disturbances in interplanetary space collectively referred to as space weather, which interacts with the Earth's upper atmosphere and causes dramatic impact on space- and ground-based technological systems [1]. Exoplanets within close-in habitable zones (HZs) around M dwarfs are exposed to extreme ionizing radiation fluxes, thus making exoplanetary space weather (ESW) effects a crucial factor of habitability [2,3]. In this paper, we describe the recent developments and provide recommendations in this interdisciplinary effort with the focus on the impacts of ESW on habitability, and the prospects for future progress in searching for signs of life in the Universe as the outcome of the NExSS workshop held in Nov 29-Dec 2, 2016, New Orleans, LA.

This is one of five "Life Beyond the Solar System" white papers submitted by NExSS. The other papers are: (1) *Exoplanet Astrophysical Properties as Context for Habitability;* (2) *Technology Development Required for Future Progress;* (3) *Remotely Detectable Biosignatures;* (4) *Observation and Modeling of Exoplanet Environments.*

**1. Areas of significant scientific or technological progress since publication of the NASA Astrobiology Strategy 2015**

From the perspective of ESW, major developments since AS15 are the following:

*A. Exoplanet Observations*

1. Discovery and characterization of superflares on K-M dwarfs, their frequency and relations to spot sizes, rotation and effective temperatures [4-6].

2. Observational search for CMEs from active stars has recently started [7,8].

3. Detection and characterization of exospheres in hot Jupiters and constraints on star-planet interaction (X-ray and Extreme UV (XUV) driven evaporation) models [9].

4. Characterization of XUV fluxes from K-M dwarfs using combined HST, EUVE, Chandra and XMM-Newton data [10,11].

5. Reconstruction of Zeeman Doppler Imaging in a number of G-M dwarfs as a prerequisite to constrain space weather models [12,13].

6. Detection of radio emission from substellar objects, extending down to a mass of 12.7 +/- 1 $M_{Jup}$, confirming magnetic field strengths >3000 G for the latter [14].

7. Development of the capability to conduct near-continuous simultaneous monitoring of 1000s of nearby systems for radio emission (stellar CMEs, planetary auroral emissions) and optical emission (stellar flares).

*B. Modeling of Stellar and Planetary Environments*

1. 3D magnetohydrodynamic (MHD) multi-fluid models of stellar winds and CMEs have recently been constructed using advanced data-driven MHD tools validated and calibrated for solar wind models [2,15-19]. These simulations suggest that fast, dense winds and powerful CMEs disturb exoplanetary magnetospheres, generate ionospheric currents, and introduce a number of effects including electron precipitation and Joule heating. **These effects need to be characterized to build a comprehensive picture of their impacts on atmospheric erosion, particularly for HZ planets orbiting M dwarfs which will be the first targets to characterize Earth-like exoplanets.**

2. 1D multi-fluid coupled hydrodynamic and kinetic models of XUV driven ion escape from exospheres of Earth-like exoplanets suggest that large XUV fluxes from active planet hosting M dwarfs stars may contribute to atmospheric erosion on geological timescales thus making exoplanets within their HZs uninhabitable [2,3]. Determining the timescales over which these stars are active and the extent of atmospheric erosion is vital for understanding exoplanet characterization and target selection with JWST.

3. 1D photo-collisional models enhanced with neutral chemistry were recently applied to model the prebiotic chemistry driven by precipitation of energetic protons due to SEPs from the young Sun and active stars [2,16,20].

*C. Technology*

1. Development of direct imaging techniques in the mid-infrared (IR) bands with Exo Life Beacon Space Telescope, ELBST (extended Fourier-Kelvin type stellar interferometers (FKSI) mid-IR space interferometers).

In the upcoming decade the exoplanet and astrobiology communities need to prepare and develop future mission concepts for space interferometry missions to directly image exoplanets in the near- and mid-IR around nearby solar type stars. The IR spectral region (3-28 microns) is well known for its richness of molecular features from bands of molecules such as carbon dioxide, water vapor, nitrous oxide, methane, hydroxyl and nitric oxide. Considerable technology development for mid-IR nulling interferometers began with the Keck Interferometer Nuller (KIN), and recently the LBTI that have provided the most sensitive observations to date of the luminosity function of warm debris disks in the HZs of nearby solar type stars. Testbeds for space interferometers (TPF-I/Darwin/FKSI) have also been developed in the US and Europe.

2. OST development.

The Origins Space Telescope (OST) is one of four mission concepts currently being studied by NASA in preparations for the Astrophysics 2020 Decadal Survey. It features a large (6.5 - 9 meter), cold (4 K), mid-to-far-IR telescope that will be orders of magnitude more powerful than existing facilities. OST will address this key science question by characterizing the atmospheres of Earth-size planets transiting in the HZs of mid-to-late M dwarf stars. OST will expand on the legacy of exoplanet science by obtaining high-precision transmission and emission (dayside and phase-resolved) spectra from 5 - 25 microns at a resolution R = 100 – 300. Achieving the necessary precision with this proven technique requires the design of a purpose-built instrument. Continued development of detector technology in the mid-IR is a fundamental step for the detection of biosignatures in exoplanetary atmospheres.

**2. Important scientific or technological topics omitted from the NASA Astrobiology Strategy 2015 and which have seen advancement since publication of the strategy**

Following the progress in our understanding space weather impacts on the Earth and Mars due to recent missions (GRACE, CHAMP, MAVEN), the exoplanetary community

[22] has initiated development of new approaches omitted from the NAS15 to characterize the impacts of ESW on close-in exoplanets around M dwarfs, including Proxima-b and TRAPPIST-1 [2,3,15,21,22].

### 3. Key research goals in the search for signs of life in the next 20 years
*A. Planet Hosting Stars:*

1. ESW models for K-M dwarfs require the following observational inputs: i. Far UV, Near-UV, XUV and radio emission fluxes; ii. Physical parameters of stellar chromospheres and coronae; iii. Surface magnetic field distribution (magnetograms).
2. Observed magnetic structures including spots and their association with flares.
3. Refine characterization of stellar ages based on a set of observables including Li, rotation, CaII H&K, patterns of magnetic activity. Thus, dedicated observations of flares on K-M stars at different phases of evolution are required along with flare frequency.

*B. Star-Planet Interactions:*

1. Develop coupled MHD, hydrodynamic and kinetic models that describe the coupling of energy flows of planet-hosting stars, and their dissipation in magnetosphere-mesosphere exoplanetary environments. This requires a well-coordinated and funded interdisciplinary effort from HAPE community.
2. Derive thresholds on parameters of space weather from stars to make a planet habitable (atmospheric neutral and ion escape rates).
3. Characterize chemistry changes due to: FUV, XUV, stellar winds, & particles.
4. Search for radio and optical stellar CME signatures by performing extended long-term observations at lower frequencies (< 10 MHz) with space or lunar radio missions.
5. Search for planetary outflows in spectral lines of H (hot Jupiters) and nitrogen and metals (terrestrial planets) driven by powerful stellar flares from active K-M dwarfs.
6. Explore when M dwarf habitable cases actually shift beyond the ice line due to severe ESW, when combined with ameliorating internal heating, including radiogenic sources as well as tidal heating within compact multi-body TRAPPIST-1 analog systems.

*C. Exoplanet Environments:*

1. Explore how ionosphere-thermosphere systems respond to extreme space weather.
2. Search for $N_2$ through mid-IR transmission and direct imaging observations, as necessary to determine how common $N_2$ is within exoplanetary atmospheres.
3. Detect the chemistry of young terrestrial-type exoplanets "pregnant" with life: signatures of prebiotic chemistry.
4. Detect signatures of hydrogen-rich (primary atmospheres) of terrestrial-type exoplanets around very young planet hosting stars.
5. Understand exoplanet magnetic dynamos, mantle activity, and the interplay between volcanic/tectonic activity and the generation of Earth-like magnetic fields.
6. Explore the role exomoons play in maintaining exoplanetary magnetic dynamos? (e.g., tidal enhancement of convection vs. the possible tidal melting of inner cores.)

### 4. Key technological challenges in astrobiology as they pertain to the search for life in extrasolar planetary systems
*A. Direct Imaging*

1. The Large Binocular Telescope Interferometer (LBTI) Hunt for Observable Signatures of Terrestrial Planets (HOSTS) study has recently set new limits for exozodi detection for solar-type stars [23]. These results demonstrate the power of LBTI for vetting potential targets for future direct imaging missions such as LUVOIR or HabEx, and the importance of completing and enlarging the study in the next few years.

2. Direct imaging techniques with FKSI-type ELBST.
Ground-based prototypes demonstrating relevant technologies and obtaining important science were the Keck Interferometer Nuller, and the LBTI HOSTS project [23]. Development of mission concepts and technologies were curtailed due to budget issues in the last decade. However, recent studies of star-planet interactions, including the interaction of coronal mass ejections with the atmospheres have shown that the atmosphere of the Earth (viewed as an, NO, and other molecules [20, 24]. Exoplanetary upper atmosphers respond strongly in the mid-IR and cools through mid-IR lines of NO and $CO_2$ and open a new potential of mid-IR spectroscopy of exoplanet atmospheres, not only with OST, but also with future ground-based and space- or moon-based nulling interferometers [25].

**5. Key scientific questions in astrobiology as they pertain to the search for life in extrasolar planetary systems**
1. How can we detect spectral signatures of prebiotically important molecules highlighting fundamental prerequisites of life including nitric oxide and nitrous oxide?
2. What chemistry of the most abundant and biologically important molecules that participate in pathways producing complex sugars, amino acids, and nucleobases can be learned from the biochemistry community studying origin of life on Earth?
3. How can astrophysics inform laboratory experiments in understanding which pathways efficiently produce biologically important molecules?
4. What steps are needed to build a unified network of theorists, observers, and laboratory scientists to explore the most efficient, laboratory validated, and calibrated methodologies to characterize the biologically important molecules with the strongest spectral signatures (high signal-to-noise, low spectral resolution) of life?
5. Can vibrant/detectable biospheres exist shielded from space weather in oceans below ice shells, beyond the classical HZ (including icy moons and nomad/rogue worlds)?

**6. Scientific advances that can be addressed by U.S. and international space missions and relevant ground-based activities in operation or in development**
1. TESS will greatly expand the population of known potentially habitable exoplanets, some of which may be selected for characterization by JWST transit transmissions spectra to look for signs of potential biosignature gases.
2. JWST will provide mid-IR transit and eclipse spectra of exoplanets around nearby stars, particularly M and K stars with exoplanets discovered by TESS, allowing characterization of their atmospheres.
3. ELTs and other ground-based platforms will greatly expand the list of rocky planets orbiting ultracool stars and characterize the atmospheres of some of them.
4. WFIRST will demonstrate the coronagraph technology for a future direct imaging mission that would study Earth-like planets, if total mission cost can be limited.

**7. How to expand partnerships (interagency, international and public/private) in furthering the study of life's origin, evolution, distribution, and future in the Universe**

1. NExSS's interdisciplinary community has an opportunity to formulate well-defined complex questions that can be addressed using a systems approach. To enhance the efficiency of this approach in searching for signs of life, we must also incorporate Origins of Life/Biology methodologies into these studies.

2. The International Space Science Institute (ISSI) is an efficient model of scientific collaboration in diverse fields of space science focusing on one fundamental challenge [26]. International science conferences are another important avenue to highlight challenges in searching for signs of life. We find that having only invited talks that set the stage for breakout discussions has been a novel approach to foster collaboration. From this perspective, the NExSS sponsored ESW workshop was a useful tool to connect and unify an emerging community that brings diverse ideas and methodologies to the table.

3. The key element of collaborative efforts should be the inclusion and coordination of international mission observations, theory, and laboratory experiments to explore laboratory validated, and calibrated methodologies to find the strongest signs of life.

4. International structures should explore observational methodologies through their national agencies with participation of public-private partnerships, such as the Breakthrough Initiative. This foundation plans to develop a low-cost mission to help search for life on Enceladus and its partnership with NASA can accelerate the project.

**References:**
1. Schrijver, C. J.; Kauristie, K. A., Alan D., Denardini, C. M. and 22 coauthors (2015) Advances in Space Res., 55, 2745 (2015)
2. Airapetian, V. S., Glocer, A., Khazanov, G. V., Loyd, R. O. P., France, K., Sojka, J., Danchi, W., Liemohn, M. W. (2017) Astrophys. J., 836L, 3A.
3. Garcia-Sage, K., Glocer, A., Drake, J. J., Gronoff, G., Cohen, O. (2017) ApJ Let, 844, L13
4. Maehara, H. et al. (2012) Nature, 485, 478.
5. Maehara, H., Notsu, Y., Notsu, S. and 5 o-authors (2017) PASP, 69, 41.
6. Davenport, J. R. A., Kipping, D. M., Sasselov, D., Matthews, J. M., Cameron, C. (2016) 821, L31.
7. Osten, R., Crosley, M. K. eprint arXiv:1711.05113.
8. Villadsen, J., Hallinan, G., Bourke, S. (2016) Proc. of the IAU Symp., vol. 320, 191.
9. Lopez, E. D. (2017), MNRAS, 472, 245.
10. Loyd, R. O. P.; France, K., Youngblood, A. and 8 co-authors (2016) ApJ, 824, 102.
11. Youngblood, A., France, K., Parke L., R. O. and 18-co-authors (2017) ApJ, 843, 27.
12. Lüftinger, T., Vidotto, A. A.; Johnstone, C. P. (2015) ASSL, 411, ISBN 978-3-319-09748-0
13. Airapetian, V. S., Jin, M., Lüftinger and 3 co-authors (2018), submitted to Nature Astronomy.
14. Kao, M. , Hallinan, G., Pineda, J. S. and 3 co-authors (2017) AAS #229, id.408.06 .
15. Cohen, O. (2017) ApJ, 835, 220.
16. Airapetian, V. S., Glocer, A., Gronoff, G., Hébrard, E., Danchi, W. (2016) Nature Geoscience, 9, 452.
17. Vidotto, A. A., Bourrier, V. (2017) MNRAS, 470, 4026.
18. Lynch, B. J., Masson, S., Li, Y., DeVore, C. R. and 3 co-authors (2016) JGR 121, 10677.
19. Dong, C., Jin, M., Lingam, M.,Airapetian, V. S., Ma, Y, van der Holst, B. (2017), PNAS, doi: 10.1073/pnas.170801011
20. Airapetian, V., Jackman, Mlynczak, M., Danchi, W., Hunt, L. (2017) Nature SREP. 7, article #14141.
21. Garraffo, C., Drake, J. J., Cohen, O., Alvarado-Gómez, J. D., Moschou, S. P. (2017) ApJ, 843, L33.
22. Airapetian, V. S. et al. (2018) Impact of ESW on Habitability, in preparation to Int. J. of Astrobiology.
23. Danchi, W., Bailey, V., Bryden, G. and 13 coauthors (2014) SPIE, vol. 9146, id. 914607.
24. Airapetian, V. S., Danchi, W. C., Chen, P. C., Rabin, D. M., Carpenter, K. G., Mlynczak, M. G. (2017) LPI Cont. No. 1989.
25. Defrére, D., Hinz, P. M., Mennesson, B. et al. (2016), ApJ, 824, 66.
26. International Space Science Institute (ISSI) http://www.issibern.ch